\documentclass{article}

\usepackage{arxiv}

\usepackage[utf8]{inputenc} 
\usepackage[T1]{fontenc}    
\usepackage{hyperref}       
\usepackage{url}            
\usepackage{booktabs}       
\usepackage{amsfonts}       
\usepackage{nicefrac}       
\usepackage{microtype}      
\usepackage{lipsum}		
\usepackage{graphicx}
\usepackage{natbib}
\usepackage{doi}
\usepackage{subfigure}

\title{Molecular Modeling of Aquaporins and Artificial Transmembrane Channels: \\ a mini-review and perspective for plants}

\author{ José Rafael Bordin$^\dagger$, Alexandre Vargas Ilha, Patrick Ruam Bredow Côrtes, \\ \textbf{ W. Silva-Oliveira, Lucas Avila Pinheiro} \\
	Departamento de Física, Instituto de Física e Matemática,\\
	Universidade Federal de Pelotas\\
	Caixa Postal 354, CEP 96001-970, Pelotas, RS, Brazil \\
	\texttt{$^\dagger$jrbordin@ufpel.edu.br} \\
	\And Elizane E. de Moraes \\
Instituto de F{\'i}sica,\\Universidade Federal da Bahia \\ Campus Universit{\'a}rio de Ondina, CEP 40210-340, Salvador, BA, Brazil\\ 
\And Tulio G. Grison, Mateus H. Köhler \\
Departamento de Física, \\Universidade Federal de Santa Maria\\ Santa Maria, CEP 97105-900, RS, Brazil}

\date{\today}

\begin{document}
\maketitle

\begin{abstract}
Aquaporins (AQPs) are a family of transmembrane channels that are found from  archaea, eubacteria, and fungi kingdoms to plants and animals. These proteins play a major role in water and small solutes transport across biological cell membranes and maintain the osmotic balance of living cells. In this sense, many works in recent years have been devoted to understanding their behavior, including in plants, where 5 major groups of AQPs have been identified, whose physiological function details still have open questions waiting for an answer. In this direction, we observed in the literature very few Molecular Modeling studies focusing on plant AQPs. It creates a gap in the proper depiction of AQPs since Molecular Simulations allow us to get information that is usually inaccessible by experiments. Likewise, many efforts have been made to create artificial nanochannels with improved properties. It has the potential to help humanity (and plants) to face water stress -- a current problem that will be worsened by Climate Change. In this short review, we will revisit and discuss important computational studies about plant aquaporins and artificial transmembrane channels. With this, we aim to show how the Molecular Modeling community can (and should) help to understand plants' AQPs properties and function and how we can create new nanotechnology-based artificial channels.
\end{abstract}

\section{\label{sec:level1} A brief introduction to aquaporin transmembrane channels in plants}

Ion channels are structures formed when specific proteins are incorporated into the phospholipid membrane (\citealt{Hille01}). The channels serve to establish an electrostatic potential gradient across the cell membrane by allowing a specific flux to pass through the membrane. There are many different ion channels in living cells. They differ in composition, pore structure, and ion selectivity (\citealt{Berneche01}). Among the many channels in biological membranes, the aquaporin (AQP) family of transmembrane proteins selectively transports water and small solutes across biological cell membranes and maintains the osmotic balance of living cells. Particularly, plants contain a large number of AQPs with different selectivity. These channels generally conduct water, but some additionally conduct physiologically important molecules such as ammonia (NH$_{3}$), carbon dioxide (CO$_{2}$), hydrogen peroxide (H$_{2}$O$_{2}$), and O$_{2} $ (\citealt{tornroth2006structural,ludewig2009plant,nyblom2009structural, khandelia2009gate, cordeiro2015molecular, chaumont2005regulation, dynowski2008molecular}).

The number of distinct channels in the AQPs is expressive since it is present in all kingdoms, including archaea, eubacteria, fungi, plants and animal (\citealt{2006biol}). For vertebrates, there are more than ten AQPs (\citealt{2022salman}), while in plants there are 5 major groups of AQPs: Plasma-Membrane Intrinsic proteins (PIPs), Tonoplast Intrinsic Proteins (TIPs), Nodulin-26 like Intrinsic Proteins ( NIPs), Small Basic Intrinsic Proteins (SIPs) and X-intrinsic proteins (XIPs) (\citealt{2006biol,2006kaldenhoff,2017romina,vajpai18}). PIPs and TIPs contribute to intracellular water balance and transcellular water flow~(\citealt{maurel08}). AQPs located in the plasma membrane, the PIPs, are mainly located in organs characterized by large fluxes of water, such as vascular tissues, guard cells, and flowers~\citealt{2018kapilan}. However, they are usually subdivided into two subgroups, PIP1 and PIP2, such that the morphology and physiological functions of the organs where the proteins are present can be completely distinct regarding to water or CO$_2$ transport~(\citealt{2006kaldenhoff}).  TIPs are also permeable to small solutes such as urea, ammonia, and hydrogen peroxide~(\citealt{dynowski2008molecular,mao15}). NIPs were initially found in the peribacteroid membrane of legume symbiotic root nodules -- nodulin proteins are expressed by plants and transferred to the membranes during the formation of the nodules~(\citealt{Fortin85}). They are permeable to numerous small molecules, including both beneficial and toxic metalloids, but usually have small water permeability~(\citealt{lopez12}). SIPs are not well characterized, and their physiological functions are not clear~(\citealt{ishi05}), while XIPs  are permeable to various solutes but only moderately to water~(\citealt{lopez13}), being evolved in osmotic regulation, H$_{2}$O$_{2}$ transport and metal homeostasis~(\citealt{noronha16}) and have been characterized in protozoa, fungi, mosses, and dicots~(\citealt{park10}).

Sure, the role and even the type of AQPs depends on the location in the plant~(\citealt{2019ding}), with AQPs being crucial for water and CO$_2$ transport. Despite that a higher water transport rate is observed inside PIP channels, there is evidence of CO$_2$ permeable AQPs in leaves in mesophyll cells and stomatal guard cells (\citealt{02terashima,2017groszmann}). Also, plants counteract variations in water supply by regulating all AQPs in the cell~(\citealt{tornroth2006structural}). In this sense, experimental results indicate that PIPs are responsible for the water transport in \textit{Arabidopsis} under hydric stress~(\citealt{2002martre}) and to control the hydrogen peroxide $(H_{2}O_{2})$ transport into plant roots (\citealt{2022israel}). Also, AQPs are evolved in the roots branching patterns in heterogeneous soil water conditions. Recently, the water flux was linked with a dynamic hormone redistribution: by closing the pores the ability of the hormone auxin to initiate lateral root development in dry soil is blocked~(\citealt{mehra22}). Also, PIPs regulate water transfer from the stigma to pollen in the early events of pollination in the crucifer family~(\citealt{ikeda97}).

Recent work on plants in the situation of oxygen deficiency, the so-called hypoxi, indicates a compensatory mechanism that increases the ability of AQPs to conduct water to compensate for the decline in hydraulic conductivity~(\citealt{2022kudoyarova}). Organs responsible for the  mechanical movement of the leaf and for the leaf hydraulic conductance also have AQPs in a key role~(\citealt{2002moshelion, lopez13}). Studies in the area of hydraulic resistance have a direct impact once water stress usually leads to damage to the crop, directly impacting plant life and crop yield, and recent findings indicate the importance of understanding AQPs to improve plantation productivity in the face of adverse situations (\citealt{2021ahmed}). O The identification of AQPs in different tissues (\citealt{2008sakurai}), the response behavior of the plant to environmental stresses (\citealt{2018kapilan,2006yu}) in addition to the crucial process of plant flowering (\citealt{2022li}) among other processes have currently been reported. One of the other several examples to be cited is about reproduction, this process in plants is related to the movement of water between cells or tissues, which action has already been observed in relationships with AQPs (\citealt{2005bots}).

Therefore, due to its importance and relevance, a great effort has been made to understand the structure and water conduction and selectivity mechanisms of AQPs in the last decades (\citealt{fu2000structure,murata2000structural,sui2001structural, de2001water,tajkhorshid2002control,tournaire2003cytosolic, hub2008mechanism,verdoucq2008structure,fischer2008ph,dynowski2008plant,aponte2010dynamics,qiu2010dynamic,nyblom2009structural,neto2016molecular, tornroth2006structural,ludewig2009plant, khandelia2009gate, cordeiro2015molecular, chaumont2005regulation, dynowski2008molecular, wang,hadidi,nikhil,yadav, hadidi2022dynamics, hub2006does, mangiatordi2016human, zwiazek2017significance}). Unfortunately, due to the insensitivity of existing experimental methods such as X-ray crystallography or NMR spectroscopy to the relatively short timeframes (in the range of nanoseconds) of water conduction events, this aim remains largely unfulfilled.  In addition, experiments at the molecular level cannot explain how water organize diffuses, and flux in a confined situation. In this context, simulations computational, the front-line approach for modeling the behavior of bio-macromolecules at the molecular level, has played an unprecedented role (\citealt{mangiatordi2016human}). Among the many possibilities that computer simulations bring, the creation of artificial nanoscience-based pores is possibly one of the most exciting.

Despite the relevance, there are few simulational works focusing on plant AQPs or in using Molecular Simulation to design new artificial nanopores to mimic, and even overcome, plants AQPs properties. In the next sections, we summarize the recent findings on Molecular Dynamics (MD) simulations of AQPs and water flow inside synthetic transmembrane channels. Next, we discuss some perspectives in this research field and how Molecular Simulations can play a significant role to increase our understanding of plant Aquaporins and lead to new technologies.

\section{Molecular dynamics simulations on aquaporins}

Molecular Simulations have emerged as a powerful tool to study the physical behavior of ﬂuids by modeling the motion and interactions of atoms and molecules. Classical MD simulations lie in a computational technique that uses Newton’s laws of motion to predict the position and momentum space trajectories of a system of classical particles.  Therefore, it is necessary to inform the intermolecular potential of particles, which are used to calculate potential energies and forces, and thus obtain the thermodynamic, transport, and structural properties of the system. This potential function accounts for all the atomic interactions, such as stretching, bending, torsional interactions, van der Waals forces, and long-range electrostatic Coulomb interactions.

An important advantage of MD is to accurately describe the complex structure of AQPs, and model the collective behavior of water in this complex structure. First, water is a complex substance to model, because of the competing eﬀects of hydrogen bonding and van der Waals interactions. In addition, the structure AQPs exist as tetrameric assemblies in which monomers act as independent water channels. Each monomer is formed by six tilted transmembrane helices (H1-H6), two short helices (HB and HE), and five interconnecting loops (LA-LE). Two characteristic domains regulate water permeation through AQPs (\citealt{de2001water,hashido2005comparative,qiu2010dynamic, cordeiro2015molecular}). One domain has two highly conserved Asn-Pro-Ala (NPA) motifs, which are held together in the middle of the membrane layer at the two short helices, HB and HE. This pathway consists of a  2nm-long pore that connects the cytoplasmic and extracellular vestibules of the protein. The formation of a narrow water file is assisted by a series of backbone carbonyl groups and hydrophilic side chains placed along the pore. At the pore center, two highly conserved NPA motifs provide selectivity against the passage of H$^{+}$ and other ions. Close to the extracellular exit of the channel, the so-called aromatic/arginine (ar/R) constriction region also contributes to selectivity. Some AQPs are permeated not only by water, but also by gasses, ions, glycerol, and other small molecules. There is a difference in the structure between mammalian and plant AQPs. In mammals, the formation of the tetrameric complex is driven solely by van der Waals interactions between the monomeric subunits. Thus mammalian AQPs are constitutively open channels and transport water whenever under osmotic stress.  In the case of plants AQPs, monomers are held together by disulfide bonds between conserved cysteine residues at the extracellular loops LA forming  are gated channels that can be regulated by various mechanisms  (\citealt{tornroth2006structural,nyblom2009structural}), for example, regulating the phosphorylation of specific serine residues  (\citealt{tornroth2006structural,ludewig2009plant, cordeiro2015molecular}). In the closed, unphosphorylated state, loop LD caps the cytoplasmic exit of the channel and blocks water passage. In this context, the water transport through AQPs across plant cells is regulated by gating mechanisms. Therefore, obtaining the crystal structure plant AQP is very complex. On the basis of these atomic structures, MD simulations have been widely employed to investigate water dynamics in AQPs, and have provided new insights into the mechanism of permeation and selectivity of AQPs (\citealt{fu2000structure,murata2000structural,sui2001structural, de2001water,tajkhorshid2002control,tournaire2003cytosolic, hashido2005comparative, hub2008mechanism,verdoucq2008structure,fischer2008ph,dynowski2008plant,aponte2010dynamics,qiu2010dynamic,tornroth2006structural,ludewig2009plant,hub2009dynamics, khandelia2009gate, cordeiro2015molecular, alishahi2018, alishahi2019novel, gravelle2014large, hadidi2022dynamics, hall2019experimental, hub2006does, Jensen2006, jensen2008dynamic, lohrasebi2019modeling, mangiatordi2016human, neumann2020molecular, ozu2013molecular, roux2004computational, wambo2017computing, wang2007molecular, zhu2004theory}).

\subsection{Molecular mechanisms of water transport}

The description at atomic resolution via MD of how water and solutes are transported in AQPS has been derived from landmark structural studies on the microbial and animal prototypes, GlpF and AQP-1 (\citealt{fu2000structure,murata2000structural,sui2001structural, de2001water,tajkhorshid2002control,tournaire2003cytosolic, hashido2005comparative, hub2008mechanism,verdoucq2008structure,fischer2008ph,dynowski2008plant,aponte2010dynamics,qiu2010dynamic,cordeiro2015molecular}). The pioneering work of Groot e Grubmüller (\citealt{de2001water}) 
described the mechanism of water permeation through the pore of an AQPS. The selectivity of this pore is established by a two-stage filter.  The first stage of the filter is located in the central part of the channel at the NPA  motifs located at the first intracellular and the third extracellular loop. Both loops are hydrophobic and dip into the membrane lining the NPA boxes directly on top of each other forming the selectivity region. The second stage is located
on the extracellular face of the channel in the aromatic/arginine (ar/R) constriction region and operates as a filter that blocks the passage of protons and other cations. These hydrophobic regions near the NPA motifs are rate-limiting water barriers and reduce interactions between water molecules (hydrogen bonds). These authors showed that water permeates in a single-file arrangement and that a fine-tuned water dipole rotation during passage is essential for water selectivity (\citealt{de2001water}). The alignment pattern was attributed to the electric fields arising from the macrodipoles of the two half helices HB and HE (\citealt{de2001water}). 

Törnroth-Horsefield et al. (\citealt{tornroth2006structural}) reported seven water molecules in the spinach leaf SoPIP2;1 protein channel via MD. The structure reveals that SoPIP2;1 is not a constitutively open channel and its conformation fluctuates between closed and open states depending on the conformation of a 20-residue cytoplasmic loop, the D-loop (\citealt{tornroth2006structural}). The authors observed the  phosphorylation of the serine residues favors the open conformation of the channel, therefore, the realization of water transport. During the simulations, this configuration  exhibited a highly correlated motion along the pore axis. In addition, a specific orientation of water molecules across the NPA motif is observed indicating the presence of a positive electrostatic potential at the NPA region to prevent proton translocation in AQPs, as reported in  of Groot e Grubmüller (\citealt{de2001water}). Later, Khandelia et al. (\citealt{khandelia2009gate}), investigated via MD  the possibility of driving the conformations equilibrium of the SoPIP2;1 protein channel toward a constitutive open state, introducing two separate mutations in the D-loop while
being in the closed conformation. The simulations suggest the permeability of the open conformation of SoPIP2;1 permeability of the open conformation $1.42 x 10^{-14}$ cm$^{3}/s$.

\begin{figure}[h!]
	\begin{center}
		\includegraphics[width=0.95\textwidth]{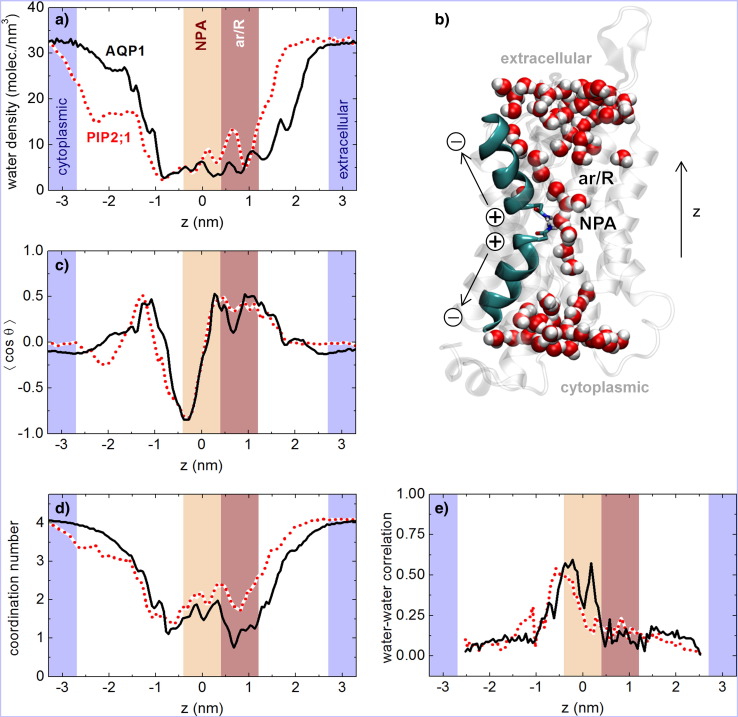}
	\end{center}
	\caption{(a) Distribution of water molecules along the pores of AQP1 (full black) and PIP2;1 (dotted red). The positions of the NPA and ar/R regions are represented in the background. (b) Simulation snapshot showing water molecules (van der Waals spheres) in the AQP1 channel (transparent grey). Helices HB and HE are highlighted in cyan with their macrodipoles as black arrows. The Asn residues at the NPA region are shown in licorice representation. (c) Orientation of water dipoles expressed as function of their angle with respect to the z-axis. (d) Coordination of water to other water molecules. (e) Correlation coefficient c(z) related to the motion of adjacent water molecules along the pore. Adapted with permission from \citealt{cordeiro2015molecular}.} 
	\label{cordeiro}
\end{figure}

Cordeiro~(\citealt{cordeiro2015molecular}) calculated via MD the distribution and organization of water molecules in AQP1 and plant PIP2;1 (Spinacia oleracia plasma membrane intrinsic protein) channels using interatomic interactions described with the GROMOS 54A7 force field for AQPS, and SPC water model (\citealt{cordeiro2015molecular}). Inside the pore, water molecules were organized as a narrow file. Their dipoles were oriented according to the electrostatic macrodipoles generated by helices HB and HE (Fig. \ref{cordeiro}a-c).  Water dipoles in different halves of the channel were oppositely oriented, meaning that molecules were forced to reorient as they passed through the NPA region. The average
coordination number to other water molecules had values between 1 and 2 due to occasional disruptions in the water file structure (Fig. \ref{cordeiro}d). In the case of AQP1, disruptions were more significant at the ar/R constriction region. The motion of water molecules close to the NPA region was highly correlated, indicating that molecules moved collectively along the pore (Fig. \ref{cordeiro}e)(\citealt{cordeiro2015molecular}).

The permeability of AQPs is based on the collective motion of
water molecules along the pore. The quantitative values of water permeability, $P{_f}$,  for the single channel osmotic permeability of mammalian AQP-1 around 3.6 ~(\citealt{cordeiro2015molecular}) 7.1~(\citealt{zhu2004theory}), 10.3 (\citealt{hashido2007water}) and $7.5 x 10^{-14}$ cm$^{3}/s$~(\citealt{de2001water}). These works showed excellent agreement with the experimentally determined 5.4 x 10$^{-14}$ cm$^{3}$/s (\citealt{engel2002aquaglyceroporins}). In the case of AQP-0 (the membrane protein of the lens fiber cells) transports water more slowly than other AQPs such as AQP-1. In the literature were reported $P{_f}$ of AQP-0 based on MD simulations are 0.2 (\citealt{hashido2005comparative,hashido2007water}) and 0.28  x 10$^{-14}$ cm$^{3}/s$ )\citealt{jensen2008dynamic}) , while reported experimental value is 0.25  x 10$^{-14}$ cm$^{3}/s$ (\citealt{qiu2010dynamic}). In the AqpZ and Glpf was related $P{_f}$ around 16$^{-14}$ cm$^{3}/s$ (\citealt{hashido2007water}) The AQPs showed a wide range of $P{_f}$ values due  that differences in the channel structures would determine the  $P{_f}$ values.

\subsection{Molecular mechanisms of physiologically molecules transport}

The physiological function of AQPs in conducting gas molecules is not fully known (\citealt{alishahi2019novel}). MD modeling identified that there are possible routes for CO$_{2}$ to cross in a membrane (\citealt{wang2007exploring}). In addition,  AQPs are capable of transporting ROS across phospholipid membranes. MD simulations (\citealt{cordeiro2015molecular}), showed that water and ROS had similar transport features and $H_{2}O_{2} $ might be transported by mammalian and plant AQPs. In addition,  recent work (\citealt{alishahi2019novel}) has shown that CO$_2$ molecules may pass through all AQP-5 monomer channels, such as the central pore. The core pore of AQP-5, like other AQPs, is known to be impervious to water molecules (\citealt{janosi2013gating}). CO$_2$ molecules, unlike water molecules, may readily pass through the AQP-5 central pore. The hydrophobic property of AQP's central pore has no influence on the passage of nonpolar CO$_2$ molecules across this channel (\citealt{alishahi2019novel}). A number of CO$_2$ molecules diffuse through the palmitoyloleyl phosphatidylethanolamine (POPE) bilayer due to the free volume in the lipid structure. The greatest energy barrier for CO$_2$ molecules in the channel for a specific CO$_2$ diffusion with velocity 1 \AA/ns is 8.8kcal/mol, but the energy barrier of the central pore does not surpass 5.1Kcal/mol for the same simulation \citealt{alishahi2019novel}. As a result, the hydrophobic central pore is the best route for CO$_2$ molecules to pass through the AQP5 (\citealt{alishahi2019novel}). 

Despite the MD simulations to quantitatively reproduce experimental parameters, even when different force field parameters and conditions were used. On the other hand, MD neglects the quantum mechanical effects of the systems. Therefore, it is impossible to consider the effects of chemical bond formation and breakage, and consequently, it was not possible to obtain proton diffusion. Furthermore, the literature lacks a greater discussion of theoretical and experimental work on water transport in open channels of AQPs plants. This is attributed to the great difficulty in obtaining the structure of these plants. 
 
\subsection{H$^{+}$ exclusion by aquaporins}

The plants AQPs membrane channels as their opening and closing gating is regulated by multiple effectors including H$^{+}$, phosphorylation, and others. Since the  determination of X-ray structures of spinach SoPIP2;1 in an open pore and a closed pore conformation, and MD of water transport in the open pore (\citealt{tornroth2006structural,khandelia2009gate}). However,the effect of phosphorylation on AQPS should be further explored in simulations. In addition, there is no experimental evidence of the efficient transport of  H$^{+}$ that any plant AQPs has, until the moment. In simulations of AQP-1 (\citealt{de2003mechanism, burykin2003really, chakrabarti2004molecular, kato2006barrier}),  showed the large electrostatic barrier impeded H$^{+}$ flux, together with the dehydration cost of moving a proton into the narrow hydrophobic channel. The electrostatic barrier excluded protons from permeating even in the presence of an intact proton wire (\citealt{chen2006origins, kato2006barrier,hub2009dynamics}).


\section{Computational studies on water flow in synthetic ionic nanochannels and membranes}

Artificial nanochannels mimicking water transport of organic membranes have been extensively investigated in the last few decades. AQP analogs are expected to help materials science and nanofluidics to establish a new era for desalination and water purification technologies (\citealt{kocsis2018}). It also means that we now have access to what extent the charge distribution, atomic arrangement, and molecular architecture are determinants for the water conduction within our very cells. There is no surprise that so many condensed matter groups worldwide are now focusing efforts on investigating the interaction at solid-liquid interfaces of solid state-based nanochannels and membranes. These artificial membranes open several new potential advantages, such as improved stability, simple and scalable fabrication, and controlled functionalization~(\citealt{Kolahalam2019,Baig2021,Speranza2021,Leo2023}) -- and no toxicity~(\citealt{daRosa2021}).

The research on synthetic aquaporin-like nanopores is focused mainly on the design and synthesis of novel conducting channels with improved water transport performances and solute rejection properties. It is important to note that a possible knowledge gap between solid-state nanochannels and naturally occurring AQPs lies in the enhanced dielectric exclusion due to the cylindrical shape of most of the artificial nanochannels being studied today (\citealt{yaroshchuk2000dielectric}). In fact, ionic rejection is very sensitive to these factors~(\citealt{Bordin12}).

Computational simulations based on classical MD or first-principles DFT have been vastly employed to decipher the details regarding water conduction in nanochannels composed of different atomic arrangements. Most recently, some groups started using artificial intelligence based on graph neural networks to enhance the search for new, improved materials to be used in desalination plants (\citealt{wang2021efficient}). Porosity and adsorption properties are also being studied with the assistance of machine and deep learning (\citealt{jian2022predicting,ogoke2022deep}).

\subsection{Water structure under subnanometer confinement}

The confinement of water in nanochannels (e.g., nanotubes and 2D materials) is able to freeze water into crystalline ice-like solids, and it can happen at ambient pressure and temperature conditions~(\citealt{kohler-pccp2017, Corti2021}). Koga et al. (\citealt{koga2000ice}) found ice structures inside carbon nanotubes (CNTs) characterized as stacked, ordered polygonal rings of water molecules. Further analysis of ice structures suggests the existence of many ice phases inside nanotubes (\citealt{takaiwa2008phase}). Over the past decades, a vast ensemble of structures has been found inside these nanotubes, making them a prominent tool for exploring water anomalies. The number, shape, and other structural properties of the confined water are extremely dependent on the size of the nanochannels.

Highly permeable water channels based on CNTs are at the forefront of translocation and separation studies. Inside their smooth structure, water behaves differently. It performs disruptive dynamics, freezing, and even multi-phases (\citealt{kotsalis2004multiphase,Bordin2014}). Recently, Farimani and Aluru (\citealt{barati2016existence}) showed the existence of multiple phases of water at CNTs under atmospheric conditions. They found vapor, high-density ice, and liquid water phases to coexist in the region within 1 nm from the surface of the nanotube. These results can explain, for example, the no-slip phenomena (\citealt{secchi2016massive}) and the fast transport of water in bundles of CNTs (\citealt{majumder2005enhanced}).

\subsection{Water flux in synthetic nanochannels}

The inner walls of a CNT are known for being electrically and mechanically smooth. It allows for the formation of a depletion layer at the interface that results in high water flux~(\citealt{kohler2018}). However, the smooth inner surface contrasts with often abrupt entrances. The problem at a usual entrance of a CNT is that much of the energy that would be converted in a highly organized stream flow is lost through friction due to collisions of water molecules and carbon atoms. Interestingly, nanotubes built from carbon atoms assembled in an AQP architecture (conical entrances with varying opening angles, see Figure 2) exhibited increased water permeability as compared to pristine CNTs, as shown in a theoretical study by Gravelle et al. (\citealt{gravelle2013optimizing}). It suggests that the hourglass shape present in a significant number of AQPs may be the result of a natural selection process toward optimal permeability. The same group further combined MD simulations with finite element (FE) calculations to explore the implications of friction and energy dissipation at the nanopore/reservoir interface (\citealt{gravelle2014large}). All to confirm that the hourglass shape inspired in AQPs enhances the water transportation efficiency down to a single-file regime, with a maximum at shallow opening angles of around 5º. This illustrates the importance of entrance dissipation in nanofluidic systems and stresses how the fine-tuning of the nanopore geometry is necessary for optimizing water transport. More recently, inspired by the work of Gravelle’s group, Hadidi and Kamali (\citealt{hadidi2020non}) have shown that an external pressure difference can lead to a water flux enhancement. A similar enhancement is observed when an external electric field is applied. The high flux regimes can be explained in terms of a reduction in the steric crowding effect of water molecules and an increase in the average number of hydrogen bonds, both occurring at the entrance of the nanochannel.

Aquaporins' capacity for mediating highly selective and fast water transport has also inspired the advance and development of different supramolecular ion-excluding artificial water channels. In a recent contribution, Roy et al.  (\citealt{roy2021foldamer}) studied different types of foldamer-derived polymer-based synthetic water channels to find that tuning of interior groups and attachment of carboxylic acid-based lipid anchors lead them to an outstanding transport of about $2.7\times10^{10}$ water molecules per second. They also found these membranes to possess a high capacity to reject both salts and protons.

\subsection{Nanopore selectivity}

The specificities of the nanopore architecture play a huge role in preventing or accelerating ionic passage. Many computational studies have been employed to understand this topic using artificial Bio-inspired nanopores. Bordin and co-authores have conducted simulations using a cylinder made by annular rings of Lennard-Jones spheres as the ionic channel~(\citealt{Bordin12}), the dielectric constant of water equals to bulk water value, K$^+$ ions with several concentrations, so they could compare their results with experiments of gramicidin A (gA). The binding sites and the selectivity in the artificial channel were obtained by inserting negative charges in two specific points of the hydrophobic nanotube structure - similar to a functionalized carbon nanotube. They found a similar current versus voltage behavior between simulations and experiments. The resemblance was verified to different concentrations. 

Among the computational alternatives available to investigate ionic transport and rejection in nanochannels, Brownian Dynamics (BD) is currently one of the most suitable tools because of its integration scheme that involves only the ionic motion. At the same time, the degrees of freedom of proteins and other bodies are kept fixed. In this method, the water is modeled as a continuum external field with a proper dielectric constant assumed (\citealt{carr2011atoms,Bordin2016}).

Many computational studies have employed CNTs to mimic organic ion channels for ion selectivity. For example, AQPs and other biological systems need selectivity between cations and anions for processes of organelle acidification or maintaining an electrochemical gradient across cells (\citealt{hille1978ionic,wang2007molecular,warshel2005inverting}).

A typical ionic exclusion and selectivity prototype involves a nanochannel small enough to offer a natural mechanical barrier (\citealt{kohler2019water,Abal2020}). Since strongly hydrated ions prefer to stay that way (preserving their electrostatic configuration), the efficiency of the membrane will depend on the hydration level (\citealt{thomas2014have}). Hydrated ions have their passage facilitated in larger pores due to an energetic penalty associated with the unbinding of water molecules from de solvation shell. However, the passage of those ions through smaller nanopores is completely hindered since they have to dissociate from water. This mechanism explains why synthetic bio-inspired membranes with nanochannels can efficiently prevent the passage of ions while allowing for unimpeded water transport. 

\section{Conclusions and Perspectives}\label{sec13}

Despite its relevance, there is a clear lack of Molecular Simulation works devoted to unraveling the details of the AQP channel's function in plants. The filling of this gap could greatly increase our understanding of the permeation of distinct molecules, such as water, CO$_2$, H$_2$O$_2$, ammonia, and urea, in PIPs and TIPs, and even shade some light in the SIPs physiological functions. Beyond that, we face a period in history where climate changes are accelerating extreme weather events, such as long periods of drought or violent storms. Plants will have to adapt or their very existence will be threatened~(\citealt{IPCC22}).

Nanotechnology and nanoscience have a central role in this urgent demand. Many studies indicate that we can insert artificial nanopores with specific functions into cellular membranes~(\citealt{hofinger11,Dutt2011,Baoukina2013, Thomas2015,Geng2014,Kanno22,Wang2022}). Then, if they can mimic the behavior of distinct transmembrane channels, why can we not improve their performance?
Recent advances in nanotube functionalization uncovered new possibilities~(\citealt{Hirsche02,Bordin17, Kharlamova2022,Hoss22,Pardehkhorram2022,Li2022,Xu2022,Turhan2022}).

Now, we can use them to control, for instance, the water permeation into plant roots under water stress by using artificial channels with distinct functionalization. We can increase or decrease the permeation to specific molecules in roots and leaves, including releasing fertilizers and nutrients at a fixed, well-defined rate~(\citealt{Yanovska2021, Vakal2022}). Even the pesticide delivery can be controlled~(\citealt{Singh2022, Wang2022a,Bhattacharya2022}), minimizing the health and environmental damages associated with the misuse of pesticides~(\citealt{Li2022a,Cara2022,Tudi2022}).

Finally, we address the fact that Molecular Modeling is a powerful tool to design nanomaterials with specific functions, providing insights that are often inaccessible in experiments. This mini-review and perspective aimed to show how urgent it is to increase the collaboration between Molecular Simulations and Plant Physiology communities, filling the gaps and increasing our understanding of molecular details of the plant machinery, and proposing ways to help to overcome the current and next challenges that plants will face, mainly due the Climate Changes.

\section*{Acknowledgments}

JRB is grateful to Gustavo M, Souza for the invitation to submit this minireview and pescpective, and for the stimulating discussions about Biology, Physics, Life and everything else.

Without public funding, this research would be impossible.  JRB is grateful to the Brazilian National Council for Scientific and Technological Development (CNPq), PQ Grant nr.  304958/2022-0, and  Research Support Foundation of the State of Rio Grande do Sul (FAPERGS), PqG Grant n2. 21/2551-0002024-5, for funding support.
MHK and JRB thanks CNPq Universal grant nr. 306709/2021-0. MHK thanks FAPERGS PqG Grant nr. 21/2551-0002023-7. LAP and TG thank CNPq and AVI, PRBC, and WSO thank the Coordination for the Improvement of Higher Education Personnel (CAPES, financing Code 001) for the Scholarship.

\section*{Declarations}

The authors disclose no conflicts of interest. All authors contribute equally to this paper.

\bibliographystyle{apalike}
\bibliography{citations.bib}


\begin{figure}[h!]
	\begin{center}
		\includegraphics[width=0.95\textwidth]{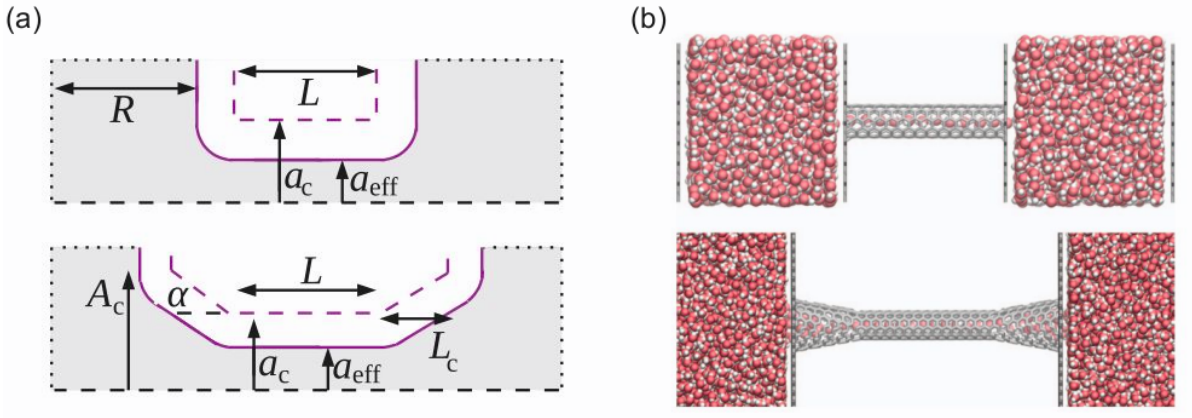}
	\end{center}
	\caption{(a) Schematic Finite Elements (FE) cylindrical and hourglass geometries and (b) the corresponding DM simulation setups. Adapted with permission from \citealt{gravelle2014large}.} 
	\label{IIIb}
\end{figure}

\end{document}